\documentclass[12pt,preprint]{aastex}

\RequirePackage{amsmath} 
\RequirePackage{amsfonts}
\RequirePackage{amssymb}

\def\ifundefined#1{\expandafter\ifx\csname#1\endcsname\relax}

\def\la{\mathrel{\hbox{\rlap{\hbox{\lower4pt\hbox{$\sim$}}}\hbox{$<$}}}}
\def\ga{\mathrel{\hbox{\rlap{\hbox{\lower4pt\hbox{$\sim$}}}\hbox{$>$}}}}

\newcommand{\be}{\begin{eqnarray}} \newcommand{\ee}{\end{eqnarray}}

\ifundefined{ensuremath}\def\ensuremath#1{\relax\ifmmode{#1}}
\else${#1}$\fi\else\relax\fi
\ifundefined{nuc}\def\nuc#1#2{\relax\ifmmode{}^{#1}{\protect\textrm{#2}}
  \else${}^{#1}$#2\fi}\else\relax\fi

\newcommand{\kmps}{\ensuremath{\mathrm{km}~\mathrm{s}^{-1}}}

\newcommand{\msol}{\ensuremath{{\mathrm{M}_\odot}}}

\def\teff{\ensuremath{T_{\mathrm{model}}}}
\def\tstd{\ensuremath{\tau_{\mathrm{std}}}}

 \newcommand{\phx}{\texttt{PHOENIX}}

\begin{document}

\title{A Self-Consistent NLTE-Spectra Synthesis Model of FeLoBAL QSOs}

\author{Darrin Casebeer \email{casebeer@nhn.ou.edu}
  E. Baron\altaffilmark{1},   Karen Leighly,  Darko
  Jevremovic\altaffilmark{2}, David Branch }

\affil{University of Oklahoma, Homer L. Dodge Department of Physics
  and Astronomy,  Norman, OK 73019, USA }

\altaffiltext{1}{Computational Research Division, Lawrence Berkeley
  National Laboratory, MS 50F-1650, 1 Cyclotron Rd, Berkeley, CA
  94720-8139 USA}

\altaffiltext{2}{Current Address: Astronomical Observatory, Volgina 7,
  11160 Belgrade, Serbia and Montenegro}

\keywords{AGN: FIRST J121442.3+280329, ISO J005645.1$-$273816}
\newcommand{\itemspace}{8.5in}
\newcommand{\FQBS}{FIRST~J121442.3+280329} 
\newcommand{\ISO}{ISO~J005645.1-273816}
\newcommand{\VR}{$\mathrm{V \propto R}$}

\begin{abstract}
  We present detailed radiative transfer spectral synthesis models for
  the Iron Low Ionization Broad Absorption Line (FeLoBAL) active
  galactic nuclei (AGN) \FQBS\ and \ISO.  Detailed NLTE spectral
  synthesis with a spherically symmetric outflow reproduces the
  observed spectra very well across a large wavelength range.  While
  exact spherical symmetry is probably not required, our model fits
  are of high quality and thus very large covering fractions are strongly
  implied by our results.  We constrain the kinetic energy and mass in
  the ejecta and discuss their implications on the accretion rate.
  Our results support the idea that FeLoBALs may be an evolutionary
  stage in the development of more ``ordinary'' QSOs.

\end{abstract}
\section{Introduction}

Spectroscopic observations of quasars show that about 10--20\% have
broad absorption troughs in their rest-frame UV spectra
\citep[see][for example]{trump06}. These absorption lines are almost exclusively
blueshifted from the rest wavelength of the associated atomic
transition, indicating the presence of an outflowing wind in our line
of sight to the nucleus. The line-of-sight velocities range from zero
to up to tens of thousands of kilometers per second
\citep[e.g.,][]{narayanan04}.

While understanding these outflows is of fundamental interest for
understanding the quasar central engine, it is also potentially
important for understanding the role of quasars in the Universe.  The
observation that the black hole mass is correlated with the velocity
dispersion of stars in the host galaxy bulge
\citep[e.g.,][]{magorrian98,fm00,gebhardt00} indicates a co-evolution
of the galaxy and its central black hole.  The close co-evolution
implies there must be feedback between the quasar and the host galaxy,
even though the sphere of gravitational influence of the black hole is
much smaller than the galaxy.  Energy arguments, however, show that is
quite feasible that the black hole can influence the galaxy; as
discussed by \citet{begelman03},  the accretion energy of the black
hole easily exceeds the binding energy of the host galaxy's bulge.

The nature of the feedback mechanism that carries the accretion energy
to the galaxy is not known.  Since AGNs are observed to release matter
and kinetic energy into their environment via outflows, it is
plausible that these outflows contribute to the feedback in an
important way. One of the difficulties in using quasar outflows in
this context is that they are sufficiently poorly understood that
there are significant uncertainties in such basic properties as the
total mass outflow rate and the total kinetic energy.


What is the kinetic luminosity of the broad absorption line quasar
winds?  That turns out to be very difficult to constrain.  While the
presence of the blueshifted absorption lines unequivocally indicates
the presence of high-velocity outflowing gas, the other fundamentally
important properties of the gas, including the density, column
density, and covering fraction are very difficult to constrain.

The density is difficult to constrain because the absorption lines are
predominately resonance transitions, and their strengths are not very
sensitive to density.  Without knowing the density, the distance of
the gas from the central engine cannot be constrained; the same
ionization state can be attained by dense gas close to the central
engine, or rare gas far from the central engine.  Density estimates
are possible when absorption lines are seen from non-resonance
transitions, but even then, they can differ enormously.  For example,
\citet{dekool01} analyzed metastable \ion{Fe}{2} absorption lines in
FBQS~0840$+$3633, and inferred a electron density $<1000-3000\rm\,
cm^{-3}$ and a distance from the central engine of several hundred pc.
In contrast, \citet{eracleous03} analyze the metastable \ion{Fe}{2}
absorption in Arp~102B with photoionization models and infer a
density of at least $10^{11}\rm \, cm^{-3}$ and a distance of less
than $7 \times 10^{16}\rm \, cm$.

The global covering fraction is also difficult to constrain directly from the
quasar spectrum; we know the gas, at least, partially covers our line of sight, but we have
little information about other lines of sight.  Covering fraction
constraints are generally made based on population statistics.  In a
seminal paper, \citet{weymann91} showed that for most BALQSOs, the
emission line properties are remarkably similar to non-BALQSOs.  Thus,
the fact that 10--20\% of quasar spectra contain broad absorption
lines is interpreted as evidence that there is a wind that covers
10--20\% of sight lines to all similar quasars, and whether or not we
see absorption lines depends on our orientation.   Alternatively, some
BAL quasars have notably different line emission than the average
quasar; examples are the low-ionization BALQSOs studied by e.g.,
\citet{boroson92b}.  These objects may instead represent an evolutionary
stage of quasars, as the quasar emerges from the cloud of gas and dust
in which it formed \citep{becker97}.

While it seems that the column density should be easy to constrain,
more recent work has shown that it can be very difficult to measure.
It was originally thought that non-black absorption troughs indicated
a relatively low column density for the absorbing gas \cite[equivalent
  hydrogen column densities of $10^{19-20}\rm \,cm^{-2}$,
  e.g.,][]{hamann98}.  But it has now been found that the non-black
troughs indicate velocity-dependent partial covering, where the
absorption covers part of the emission region, and the uncovered part
fills in the trough partially \citep[e.g.,][]{arav99}.  Thus, the
column density appears to be high, but it is very difficult to
constrain directly from the data except in a few very specialized
cases \citep[see for example][]{gabel06,arav05}.

How can we make progress on this problem?  It is becoming clear that
because of the difficulties described above, the traditional
techniques for analysis of troughs (e.g., curve of growth) and
modeling (e.g., photoionization modeling to produce absorption line
ratios and equivalent widths) are limited.  An approach that may be
profitable is to construct a physical model for the outflow, and
constrain the parameters of the model using the data.

Our first foray into constructing physical models for quasar winds was
performed by \citet{branch02}.  In that paper, the
FeLoBAL\footnote{FeLoBALs are distinguished by the presence of
  absorption in low-ionization lines such as \ion{Al}{3} and
  \ion{Mg}{2} as well as absorption by excited states of \ion{Fe}{2}
  and \ion{Fe}{3}.} FIRST~J121442$+$280329 was modeled using SYNOW, a
parameterized, spherically-symmetric, resonant-scattering, synthetic
spectrum code more typically used to model supernovae
\citep{fisher00}.  The difference between this treatment and a more
typical one applied to the same data by \citet{dekool02b} is that
SYNOW assumes that emission and absorption are produced in the same
outflowing gas.  In contrast, the approach taken by \citet{dekool02b}
assumes that absorption is imprinted upon a typical continuum+emission
line quasar spectrum; that is, the absorbing gas is separated from the
emission-line region. In fact, based on the analysis of the
\ion{Fe}{2} metastable absorption lines, they find that the  absorber
is 1--30 parsecs from the central engine, much farther than the quasar
broad emission-line region. Note that FIRST~J121442$+$280329 is not
the only object that can be  modeled using SYNOW; \citet{casebeer04}
present a SYNOW model of another FeLoBAL, ISO~J005645.1$-$273816.

The SYNOW model is attractive because it is simple; only one component
is needed to model both the emission and absorption lines.  However,
this model is limited.  The primary purpose of the SYNOW program is to
identify lines in complicated supernova spectra.  Thus, individual
ions can be added to a SYNOW run at will in order to see if features
from emission and absorption from those ions is present.  It does not
solve the physics of the gas, so physical parameters beyond the
existence of a particular species and its velocity extent cannot be
extracted from the results.

We test the ideas of \citet{branch02} and \citet{casebeer04} by using
the generalized stellar atmosphere code \texttt{PHOENIX} to model the
spectra of the two FeLoBALs that were successfully modeled using
SYNOW, and including spectra that extend to rest-frame optical
wavelengths for FIRST~J121442$+$280329.  {\tt PHOENIX} is a much
different code than \texttt{SYNOW} in that it contains all the
relevant physics to determine the spectrum of outflowing gas.  It
solves the fully relativistic NLTE radiative transfer problem
including the effects of both lines and continua in moving flows.  For
a discussion of the use of both \texttt{SYNOW} and \texttt{PHOENIX} in
the context of modeling supernovae spectra, see \citet*{bbj03}.  We
find that \texttt{PHOENIX} is able to model the spectra from these
objects surprisingly well, and we are able to derive several important
physical parameters from the model.

In \S  \ref{Models} we describe the \texttt{PHOENIX} model in detail.
In \S  \ref{Modelfits} we describe our determination of the
best-fitting model.  In \S  \ref{Results} we describe the results of
our model fitting.  In \S  \ref{Discussion} we discuss the physical
implications of the model, how it   relates to other BAL spectra and
where it fits in the BAL picture.  An appendix 
includes a flowchart of a \phx\ calculation.

\section{Models \label{Models}}

Photoionization codes have been essential in understanding emission
and absorption features in the spectra of active galaxies.  {\it
Cloudy} \citep{ferland98,ferland03} is widely used.
\phx\ solves the
radiative transfer equations exactly, and tries to have the most
accurate radiative data possible.  As a radiative transfer code, \phx\
correctly produces line profiles due to relativistic differential
expansion. It is this feature that we are making use of here.

In this paper we model the situation envisioned by \citet{branch02}:
the emission and absorption occur in a fairly optically thick
expanding shell illuminated from the inside by the continuum; as
discussed in \S 5.4, this situation may be a consequence of quasar
evolution, occurring when the quasar ejects a shroud of dust and gas
\citep[e.g.,][]{voit93}.   

\phx\ is a spectral synthesis
code; the direct output is a model spectrum.  The only way to obtain
fluxes or equivalent widths of lines in a \phx\ model is to measure
them directly from the synthetic spectrum in the same way that they
are measured from the observed spectrum.  Measuring emission and
absorption lines from complex quasar spectra is well known to be
rather uncertain, as a consequence of blending and uncertain placement
of the continuum.  So in \phx, this step is bypassed, and the
synthetic spectrum is compared directly with the observed spectrum.
Second, the input parameters are somewhat different.  
In \phx, density is given as a function of the radius in
concordance with the assumed velocity profile as a function of radius.
An analogy to the photoionizing flux is a little difficult to
construct.   As noted in the next section, two of the important
parameters are the reference radius $R_0$, the radius at which
the continuum optical depth at 5000\AA\ is unity, and the model
temperature $T_{model}$, defined in terms of the total
bolometric luminosity in the observer's frame,
$L$, and the reference radius.  Thus, $L$ or $T$ are
somewhat analogous to the photoionizing flux, because for a fixed
reference radius, they give the intensity of the continuum at the
reference radius.  Finally, the column density can be evaluated for
particular values of the optical depth.

\subsection{The Model Parameters} 

Our models are spherically symmetric, with homologous expansion ($v
\propto r$).  Homologous expansion is analogous to the Hubble
expansion.  The model atmospheres are characterized by the following
parameters \citep[see][for details]{bsn99em04}: (i) the reference
radius $R_0$, the radius at which the continuum optical depth in
extinction ($\tstd$) at 5000\AA\ is unity; (ii) the model temperature
$\teff$, defined by the luminosity, L and the
reference radius, $R_0$, [$\teff=(L/(4\pi R_0^2 \sigma))^{1/4}$],
where $\sigma$ is Stefan's constant; (iii) the density structure
parameter $v_e$, [$\rho(v) \propto e^{-v/v_e)}$]; (iv) the expansion
velocity, $v_0$, at the reference radius;  (v) the pressure, $P_{out}$,
at the outer edge of the atmosphere; (vi) the LTE-line threshold
ratio, equal to $5\times 10^{-6}$; (vii) the albedo for line
scattering (metal lines only, here set to 0.95); (viii) the
statistical velocity $\zeta=50$~\kmps, treated as depth-independent
isotropic microturbulence, and (ix) the elemental abundances, assumed
to be solar as
given by \citet{gn93}.

We emphasize that for extended model atmospheres one should not
assign, \emph{a priori}, a physical interpretation to the parameter
combination of \teff\ and $R_0$. While \teff\ has a well defined
physical meaning for plane-parallel stellar atmospheres, its
definition for extended atmospheres is connected to the particular
definition of the radius $R_0$ \citep[see][]{bsw91}. In addition, the
reference radius $R_0$ in our models is defined using an extinction
optical depth scale at $\lambda=5000$\AA\ and is not directly
comparable to observationally derived radii. Therefore the model
temperature is not well defined for extended atmospheres and must be
regarded only as a convenient numerical parameter.  We chose the
maximum extinction optical depth so that model would be just optically
thick to continuum scattering in order to replicate the optical
spectrum observed in these objects (see \S~\ref{results:fqbs}).

\subsection{Comparison of our model with \citet{dekool02b}}

\subsubsection{The \citeauthor{dekool02b} approach}

Because we want to compare and contrast our model with
\citet{dekool02b} we must first briefly describe their analysis.  We
repeat the description of their analysis from   \citet{branch02}
here. In their analysis, \citet{dekool02b} used effective continuous
spectra that consisted of a power-law continuum plus \ion{Fe}{2} and
\ion{Mg}{2} Broad Emission Lines (BELs).  The \ion{Mg}{2} BEL was the
sum of two Gaussians 
centered on the two components of the \ion{Mg}{2}
$\lambda$$2798$\AA\ doublet $(\lambda$$2796$\AA, $\lambda$$2803\AA)$.
Two different templates for the \ion{Fe} {2} BELs were considered.
The first consisted of a linear combination of five sets of
\ion{Fe}{2}  BELs from theoretical model calculations (Verner et al.
1999), and the second was the observed \ion{Fe}{2} BEL spectrum of the
strong emission-line QSO 2226--3905 \citep{graham96}.  For the
absorption features, \citet{dekool02b} obtained a template
distribution of line optical depth with respect to velocity in the
Broad Absorption Line Region (BALR) from the observed absorption
profile of \ion{Fe}{2} 
$\lambda$3004\AA, an apparently unblended line of moderate
strength. Given the assumption that only absorption takes place in the
BALR, the optical depth was obtained from $\tau$(v) = $\log$
F$_\lambda$, where F$_\lambda$ is the fractional  residual flux in the
absorption feature. The resulting optical depth distribution extended
from about 1200 to 2700 \kmps\ and peaked near 2100 \kmps. This optical
depth distribution, scaled in amplitude, was used for all absorption
lines. For each of the absorbing ions that were introduced
\ion{Fe}{2}, \ion{Mg}{2}, \ion{Cr}{2}, and \ion{Mn}{2} the column
density was a fitting parameter. The relative strengths of the lines
of each ion turned out to be consistent with LTE.

Three models that differed in their details \citep{dekool02b} were
presented, with similar results. The column densities of \ion{Fe}{2},
\ion{Cr}{2}, and \ion{Mn}{2} were well constrained. (The column
density of \ion{Mg}{2} could not be well constrained because the only
\ion{Mg}{2} absorption, due to $\lambda$2798, is saturated.) The
excitation temperature was found to be near 10,000 K. Two local
covering factors, representing the fractions of the power-law source
and the BELR that are covered by the BALR as viewed by the observer,
were introduced to reproduce the observed nonblack saturation, that is
the fact
that in the observed spectrum, even very strong absorption features do
not go to zero flux. Both local covering factors were found to be $0.7
\pm 0.1$.

A detailed view of the spectral fit for one of the models was
presented, and practically all of the observed absorptions were
reasonably well accounted for. In order  to interpret the results of
\FQBS.  of their spectrum fits, \citet{dekool02b}  used the
photoionization--equilibrium code \emph{Cloudy}  to compute a grid of
constant-density slab models irradiated by a range of ionizing
spectra. The ionization parameter $U$, the hydrogen density $n$ and
the hydrogen column density $N_H$ were found to satisfy  $2.0 < \log U <
0.7$, $7.5 < \log n < 9.5$, and $21.4 < \log N_H <  22.2$,
respectively. From these values, the distance of the BALR from the
center of the QSO was inferred to be between 1 and 30 pc.

\subsubsection{The PHOENIX approach}

\texttt{PHOENIX} calculations are very detailed,  starting from
first principles. We compare it with the rather simplified, template
fitting approach of \citet{dekool02b} \citep[see
also:][]{korista92,arav01,dekool01} and discuss differences in the
approaches. More details of \texttt{PHOENIX} are presented in an 
Appendix.

In contrast to the approach by \citet{dekool02b} we model the
continuum, the emission lines, and the absorption lines
simultaneously.  The \ion{Fe}{2} emission lines in our model are
calculated self-consistently using our rather large model
atom. The velocity width of the emission
lines is determined by the 
outflow characteristics.  In fact, the radiative transfer through the
moving clouds is what gives the lines their width and therefore the
calculations from the model are directly comparable with the observed
spectrum. In addition we have run in excess of 100 models during the
fitting process fitting the continuum, emission and absorption line
features simultaneously, compared with the linear combination of the
five calculations run with the sophisticated \emph{Cloudy} models in the
\citet{dekool02b} approach.  None of our models are based on observed
emission line spectra and as such we do not have the problems
associated with the correct placement and removal of the continuum.

We have none of the fitting parameters associated with the emission
line fits displayed in
\citet{dekool02b}. We rely on a global fit to the overall spectrum
from a model calculated from first principles, to determine how our
model compares with observation.

Our approach does not use a
template fit to the absorption
line spectrum $\tau$(v) = $\log$ F$_\lambda$.  In our model the
spectrum is calculated directly from self-consistently solving the
radiative transfer problem with scattering in an expanding atmosphere
from first principles. The full spectral energy distribution,
including continua, emission, absorption, and the effects of
differential expansion are solved for simultaneously and then the
results are compared to the observed spectra. Our optical depth
profile as a function of 
velocity is determined based on the physical conditions of the gas and
the radiation field at a given velocity, not on the observed spectrum of
the object \emph{a priori}.

Our model grid is specified in terms of \tstd\ and the spatial extent
is then determined by the implicit condition that \tstd($R_0$)$ =
1.0$. Thus, the inner and outer spatial boundaries are determined by
the input \tstd\ grid and vary not only from model to model but within
a given model from iteration to iteration as we reach
convergence. Given the density parameterization, the pressure and
temperature are determined by the NLTE equation of state from
iteration to iteration. The models are very well converged. In
contradistinction to the template fitting of method \citet{dekool02b},
since we do 
not have a spatially fixed inner boundary and the \tstd\ grid is fixed,
we do not have \emph{a priori}, a fixed ionization parameter or column
density.

\section{Determining the Best Fit Model\label{Modelfits}}

In order to find  the appropriate model parameters we began with the
parameters of \citet{branch02} 
and restricted our initial calculations
to pure LTE, which is much less computationally demanding and hence
allows us to produce large grids which are then calculated in full
NLTE.  Given an LTE grid, we then chose a subset of ``best fit'' (by
eye) models which we proceeded to calculate in full NLTE.  Once we
felt subjectively that we had the best model fit we turned to a more
objective method of determining the best fit.  Over 100 models were
run during this process.

\begin{figure}
\centering
\includegraphics[width=0.8\textwidth,angle=0]{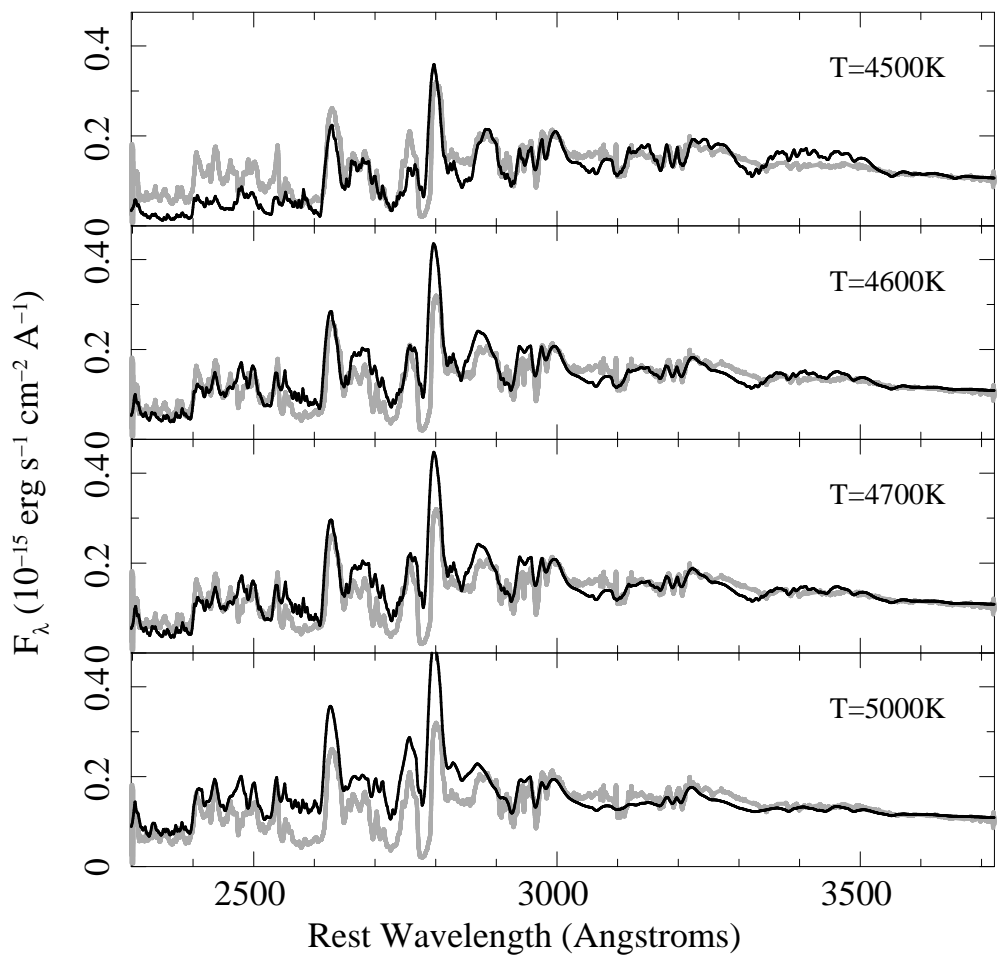}
\caption{\label{fig:show_fit_proc} Four models (black lines) are
  shown, normalized to the observed spectrum of FIRST J1214+2803 (grey
  lines) at 3600\AA.  The top and bottom models are seen to provide a
  distinctly poorer fit than the middle two models, especially in the
  2300--2900\AA\ band.  The $T=4600\rm \,K$ and $T=4700\rm \,K$
  provide a much better fit, and although they appear very similar,
  the $T=4600\rm \,K$ was selected as the best fit by minimizing a
  $\chi^2$-like figure of merit.}
\end{figure}

Determining a goodness of fit criterion for spectra is a non-trivial
task. A pure $\chi^2$ method is not ideal because a good fit of a
synthetic spectrum to an observed spectrum should ideally fit on all
scales. That is, the continuum shape (colors) should be correct as
well as all the line features. Since our synthetic spectral models
include detailed physics of the interaction of the lines and continua,
we strive to fit both. Since we don't know the errors due to flux
calibration, reddening, etc., a $\chi^2$ would not have the
traditional meaning in terms of
probability. Figure~\ref{fig:show_fit_proc} illustrates how the
direct comparison of models to observations clearly leads to the best
fit spectrum. While the two models with \teff = 4600 and \teff = 4700
look quite close, we used a $\chi^2$-like test found the smallest
distance between the model and the observed treating the normalization
as a nuisance parameter.


\section{Results\label{Results}}

\subsection{\FQBS\label{results:fqbs}}

We first turn our attention to the FeLoBAL quasar
\FQBS\ \citep{BeckerFBQS00,White} UV and optical spectra. We adopted a
redshift for \FQBS\ of $z=0.692$ as measured from the peak of the
\ion{Mg}{2} emission feature.  We corrected for a Galactic extinction
of $E(B-V)=0.023$ \citep{ExtEBminV} obtained from NED using the
reddening curve of \citet{cardelli89} supplemented by the work of
\citet{odonnell94}.  We smoothed all observed spectra using a
near-Gaussian smoothing procedure with a width of 300 \kmps.

\begin{figure}
\includegraphics[height=0.9\textheight]{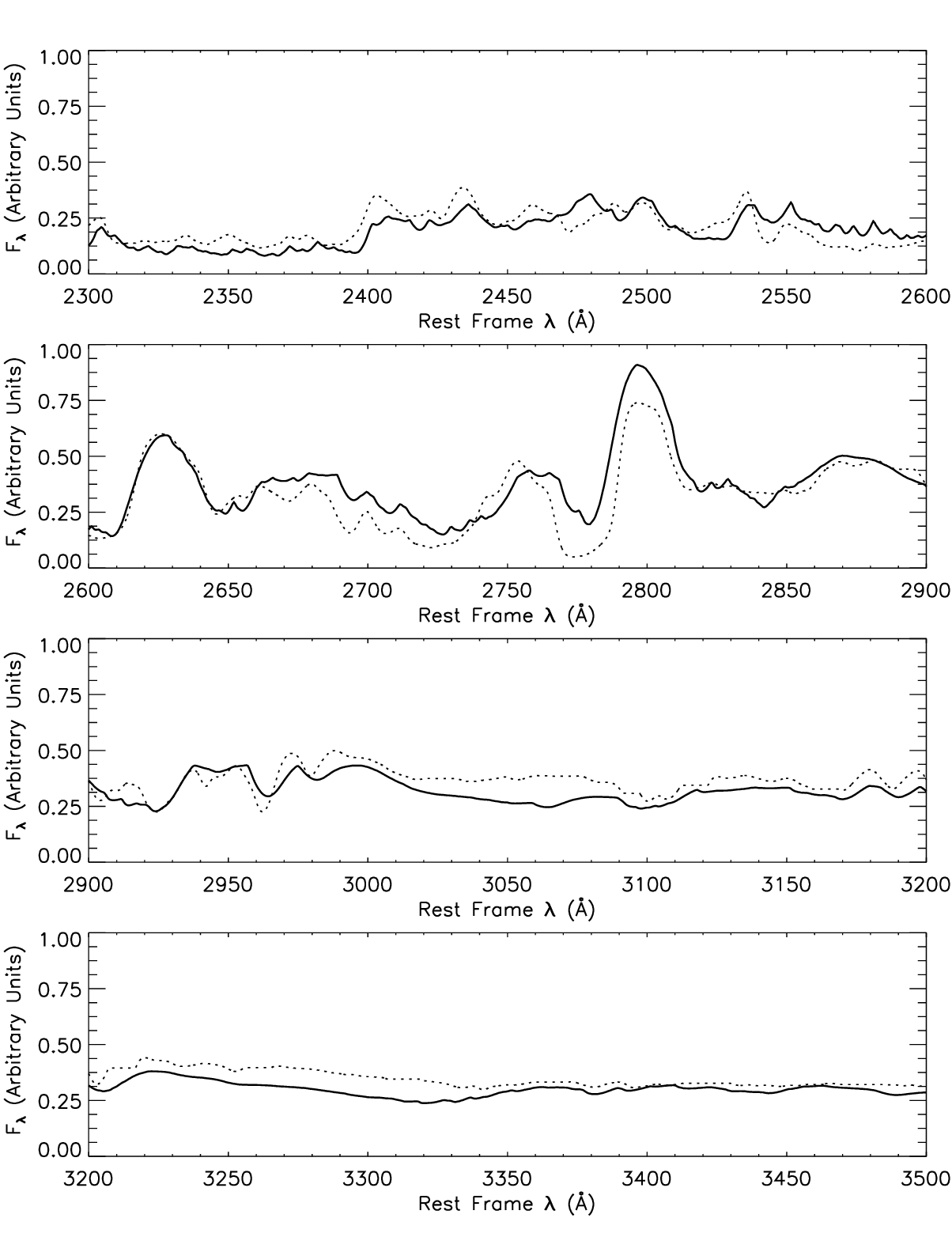}
\caption{\label{FQBSUV} The \phx\ spectrum (solid line) compared with the
  restframe, dereddened, and smoothed, observed \FQBS\ (dotted 
  line)  UV spectrum. }
\end{figure}

The model calculations and fits were done according to the method
discussed  in \S~\ref{Modelfits}. Figure~\ref{FQBSUV} compares our
best-fitting synthetic \phx\ spectrum and the restframe UV spectrum of
\FQBS.  The best-fit parameters for \FQBS\ were: $\teff=4600$~K,
$R_0=1.4\times10^{17}$~cm, $v_e=300$~\kmps, and $v_{0}=2100$~\kmps
(for completeness these values are also shown in Table \ref{BESTFITTING}).



\begin{figure}
\includegraphics{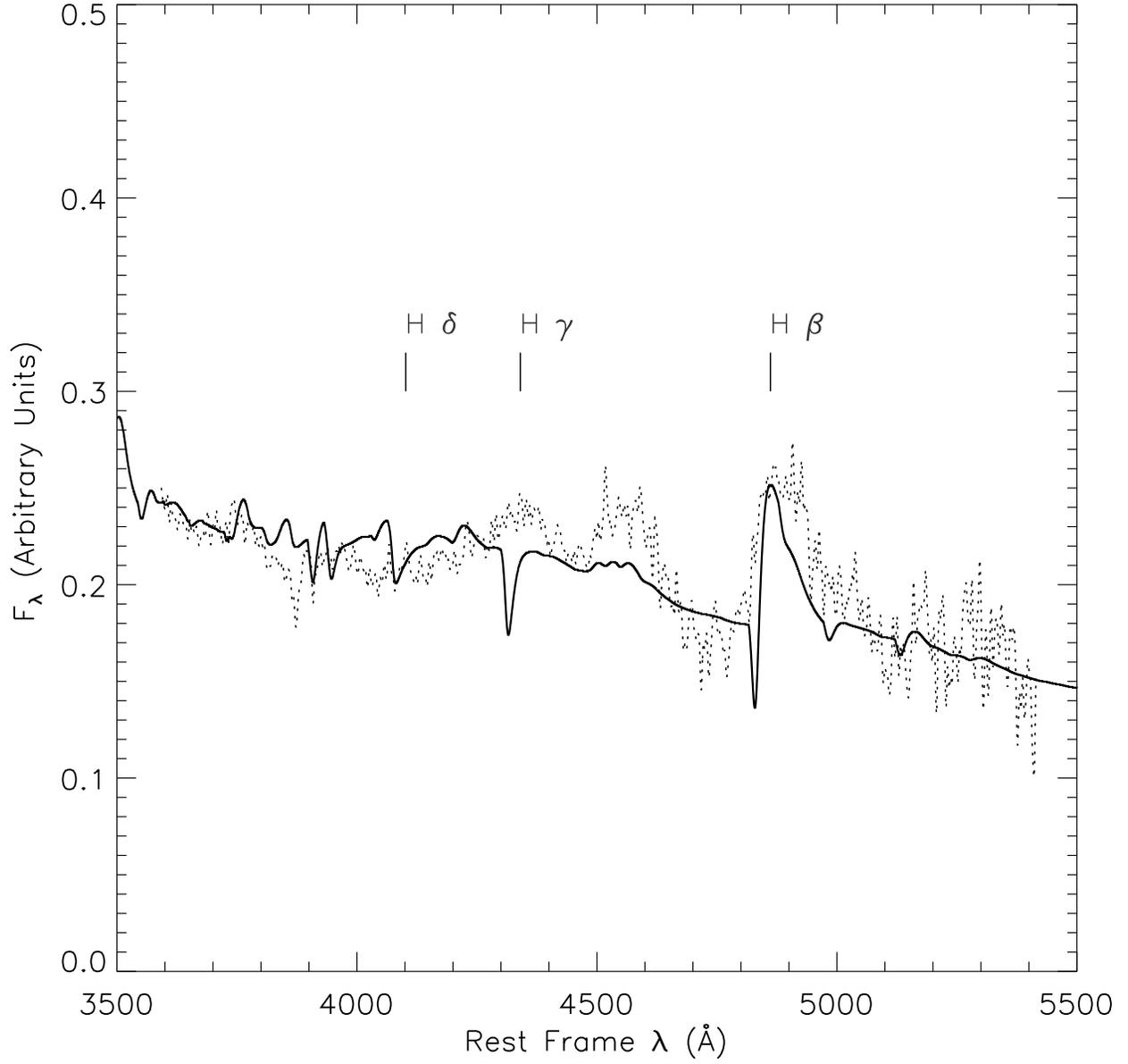}
\caption{\label{FQBSOPT} The \phx\ model (solid line) compared with the
  restframe, dereddened, and smoothed, observed optical spectrum of
  \FQBS\ (dotted line).  }
\end{figure}

With this radius and model temperature, the luminosity of the model is
$L=6.3\times10^{45}$~\ ergs~s$^{-1}$.  Overall, the synthetic model
compares favorably with the restframe UV observation of \FQBS; in
particular the \ion{Fe}{2} lines, both from ground and excited states,
fit very well. The \ion{Mg}{2} $\lambda 2798$ feature appears to be
too strong in emission and yet too weak in absorption. This could be a
sign of asymmetry in the atmosphere, \citet{branch02} and
\citet{dekool02b} found a similar result in their
calculations.  Figure~\ref{FQBSOPT} compares the \FQBS\
rest-frame optical observations to the synthetic model.  The synthetic
model calculation is a reasonable fit to the optical observation. The
features at $4600$\AA\ are reproduced in the synthetic spectrum, yet
are not as strong as observed and \ion{H}{1} absorption appears
to be too strong in the synthetic spectrum.  However, this is dependent on
our placement of the continuum and it should be noted that the
observed optical spectrum is very noisy.  The observations of
\citet{hall07} show that Balmer absorption features do exist in optical
spectra of one FeLoBAL QSO which is very similar to the QSOs studied
here.

In order to reproduce the optical observation with our model we placed
our lower boundary condition (that the specific intensity is given by
the solution to the diffusion equation at the inner edge of our opaque
core) at the somewhat low value of $\tstd = 10$. Ideally we place that
opaque core at $\tstd \ga 100$, but when we did that we found that the
model spectrum longward of 2800 \AA\ no longer  fit the
observation. In particular the optical flux was very attenuated and
had deep P-Cygni profiles unlike the observation.
 FeLoBALs, although very
optically thick by  AGN standards, are therefore thought to be
optically thinner than, for 
example, the atmosphere of a Type II supernova. Models in which the
continuum optical depth is high ($\tau_{std}=100$) fail to reproduce
the optical, and the UV spectra redward of Mg II of FeLoBALs. This
could be an indication that our inner boundary condition would be
better replaced by an AGN-like continuum, but that is beyond the scope
of the present work.

\begin{deluxetable}{lllll}
\tablewidth{0pt} \tablecaption{\label{BESTFITTING} PHOENIX
  best-fitting model parameters.}  \tablehead{
  \colhead{$R_0$}&\colhead{$v_{0}$}&\colhead{$v_e$}&\colhead{log($L_{bol}$)}}
\startdata
$1.4\times10^{17}$~cm&$2100$~\kmps&$300$~\kmps&$45.8$~\
ergs~s$^{-1}$\\

\enddata
\end{deluxetable}

\subsubsection{Importance of NLTE effects}
\begin{figure}
\includegraphics{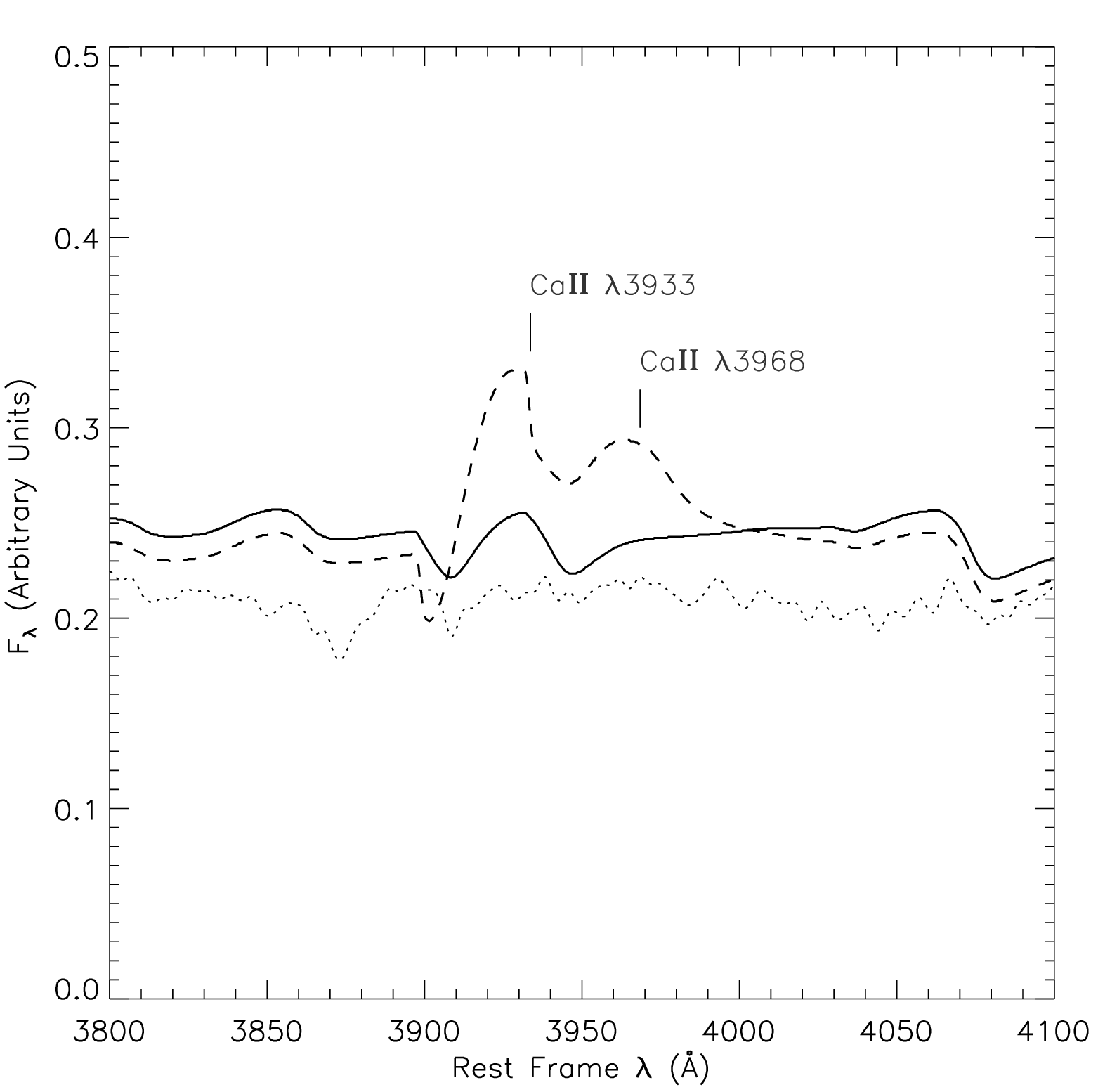}
\caption{\label{FQBSCaII} The optical spectrum of \FQBS\ (dotted)
  deredshifted, dereddened, and smoothed, versus two synthetic
  \texttt{PHOENIX} models.  The dashed line represents a
  \phx\ model with the ions discussed in \S \ref{Models} in NLTE
  except for the Calcium I-III ions, which are in LTE. The solid
  line shows a model with all of the ions in NLTE that are discussed
  in \S \ref{Models}. The fact that the synthetic spectrum matches the
observed spectrum significantly better indicates the importance of
NLTE in the calculations.}
\end{figure}

It is important to correctly model
photoionization and recombination in the models.
Figure~\ref{FQBSCaII} shows the importance of NLTE effects in
correctly modeling spectra. The solid line has all the species that we
included in these calculations in NLTE, whereas the dashed line has
everything in NLTE except for Ca I-III.  For species treated in LTE
the Saha-Boltzmann equations are solved in order to calculate the
atomic level populations instead of the full rate equations.  An LTE
treatment of calcium  results in  under-estimation of the ionization
of calcium in the model which creates a large quantity of
\ion{Ca}{2} in the line forming region. Therefore the \ion{Ca}{2} H\&K
features appear in the LTE spectrum. When NLTE is turned on for Ca
I--III in the PHOENIX model, the level populations are controlled by
the hotter radiation field (photoionized) from deeper layers and 
calcium is overionized compared to the local gas temperature.  Therefore the
\ion{Ca}{2} H\&K features, which are resonance transitions, disappear
from the synthetic spectrum.   The synthetic model which treats Ca
I--III in NLTE closely reproduces the optical observation whereas the
synthetic model without the NLTE Ca I--III clearly overestimates the
strength of the H\&K lines.

\subsection{\ISO}

The optical spectrum of \ISO\ was obtained in September 2000 with the
FORS1 instrument installed on the VLT UT1/Antu \citep{Duc}.  We
adopted a redshift of $z=1.776$ which was determined by the
\ion{C}{4}, \ion{Fe}{2} $\lambda 2627$ line, and \ion{Mg}{2} emission
lines.  We corrected for an Galactic extinction of $E(B-V)=0.017$
\citep{ExtEBminV} using the standard reddening curve
\citep{cardelli89,odonnell94}.  We smoothed all observed spectra using
a  near-Gaussian smoothing procedure with a width of 300 \kmps.

\begin{figure}
\includegraphics[height=0.9\textheight]{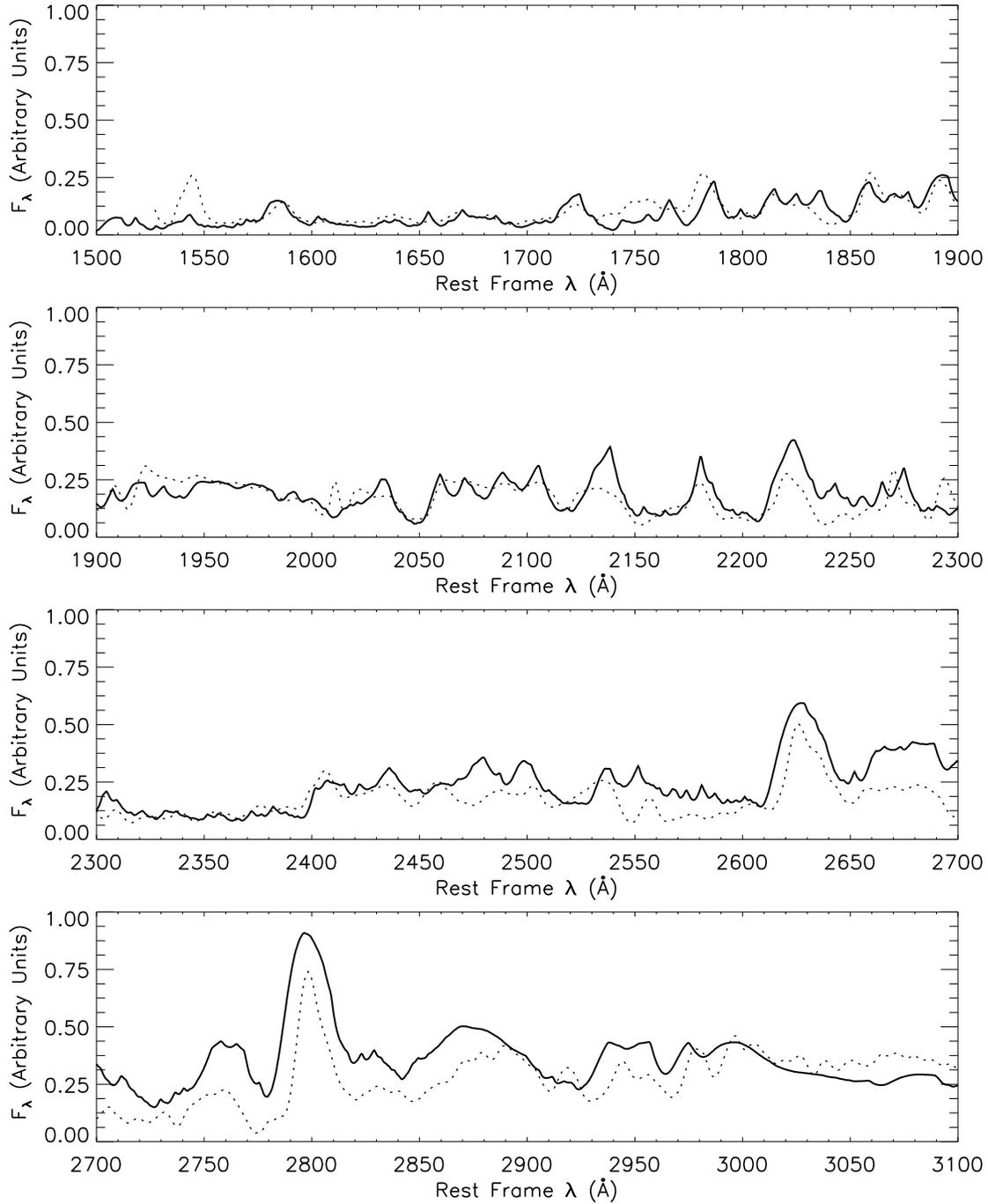}
\caption{\label{ISOUV} The \phx\ model (solid line) vs  the 
  restframe, dereddened, and smoothed,  UV spectrum of \ISO\ (dotted
  line).}
\end{figure}

The \ion{Mg}{2} feature has a similar shape when compared with
\FQBS\ and \ISO\ appears to have similar \ion{Fe}{2} features when
compared with \FQBS.  Because of the similarities between \ISO\ and
\FQBS\ we compare the synthetic model spectrum which fit \FQBS\ with
the rest frame UV spectrum of \ISO.  This comparison is shown in
Figure \ref{ISOUV}.  Following the same fitting procedure outlined in
\S~\ref{Modelfits} and the same grid of models used for \FQBS\, we
found that the parameters which were a best fit for \FQBS\ also were a
best fit for \ISO.  The \ion{Fe}{2} lines fit very well for this
object and the \ion{Mg}{2} emission feature was too strong while the
absorption was too weak. This is very similar to the fit to the UV
spectrum of \FQBS.

\begin{figure}
\includegraphics{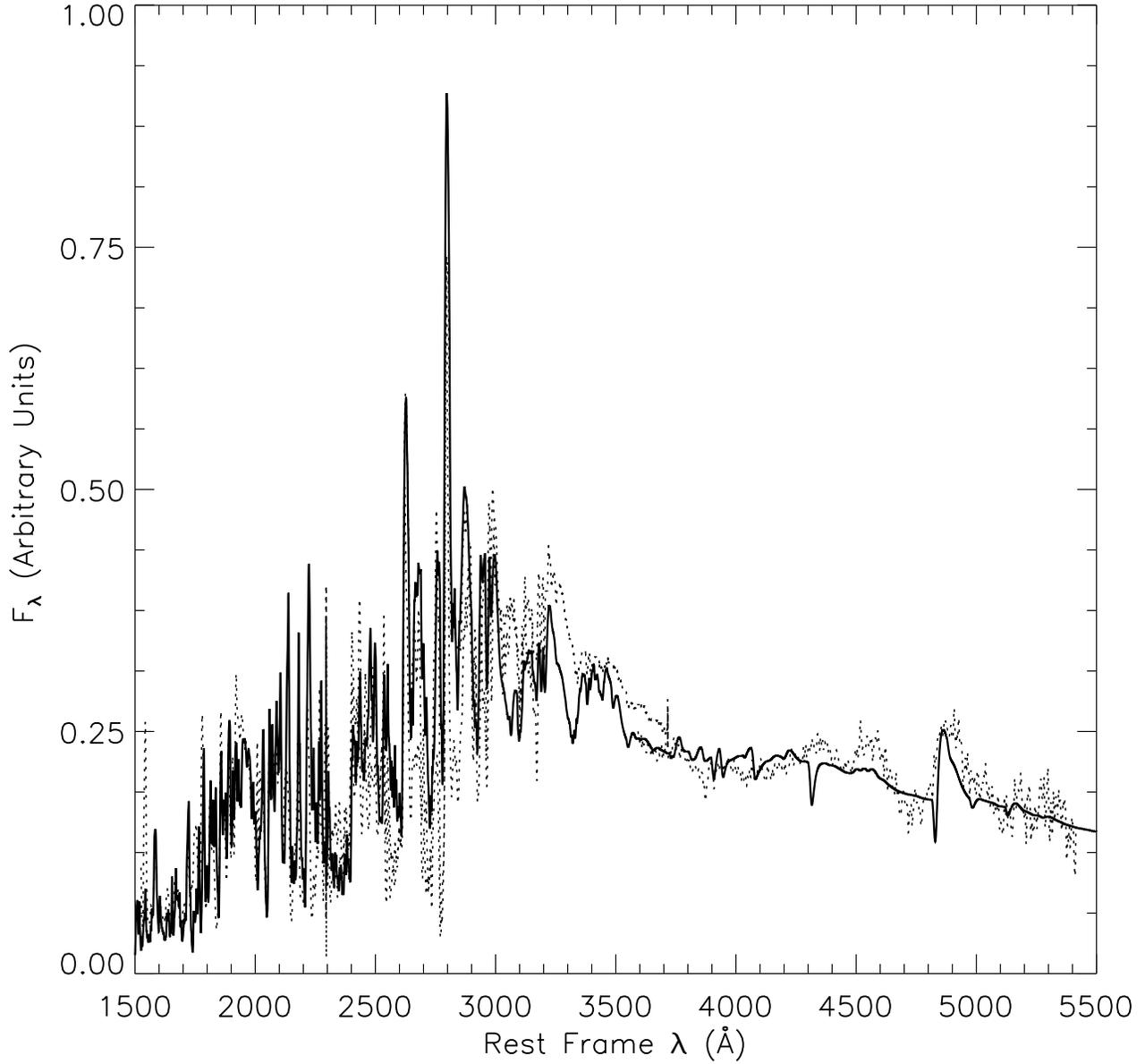}
\caption{\label{FQBSISO} The combined spectra of \FQBS\ and \ISO.  The
  wavelength range spans 1500--5500\AA.  the \phx\ spectrum is the
  solid line, while the \FQBS\ UV through optical spectrum and
  the \ISO\ UV spectrum are dotted.   All spectra are restframe,
  deredshifted, and smoothed.  }
\end{figure}

The similarities between the UV spectra of \FQBS\ and \ISO\ are
interesting. \FQBS\ and \ISO\ may be a subtype of FeLoBAL AGN with
very similar characteristics.   For emphasis, in Figure \ref{FQBSISO}
we show the synthetic model spectrum with the combined \FQBS\ UV and
optical spectra and the \ISO\ UV spectrum. The high quality fit, over
a wide wavelength range, is compelling.  It is however unlikely that
the \phx\ models would compare favorably with the total composite
spectrum of \ISO. \ISO\ was discovered in the infrared and the UV
spectrum is diminished with respect to the infrared \citep{Duc};
however the IR emission could be enhanced by  dust emission in a
physically separate region.

\subsection{Physical Conditions \label{PHYSCOND}}

The physical conditions in atmosphere are calculated from the solution of
the radiative transfer equation, the equilibrium rate equations, and
the equation of radiative equilibrium for the best fit 
input parameters given in Table \ref{BESTFITTING}.
The \phx\ model for \ISO\ and \FQBS\ has the following physical
dynamics: outflow mass $550\ M_{\odot}$, kinetic energy
$30\times10^{51}$ ergs, and a mass loss rate of
$\dot{M}=159$~\msol~yr$^{-1}$ above the ``photosphere''
($\tau_{std}=1$).  In addition the \phx\ model yields an outflow mass of
$3000\ M_{\odot}$, kinetic energy of $100\times10^{51}$ ergs,  and a mass
loss rate of $\dot{M}=466$~\msol~yr$^{-1}$ above $\tau_{std}=10$.  The
\phx\  model has  an equivalent hydrogen column density of
$2\times10^{24}$\ cm$^{-2}$ for the region above the ``photosphere''
($\tau_{std}=1$) and a maximum equivalent hydrogen column density of
$2\times10^{25}$\ cm$^{-2}$ for the  entire atmosphere
($\tau_{std}=10$). These values are displayed in Table~\ref{PHYSCONDTAB}.

\begin{deluxetable}{lllll}
\tablewidth{0pt} \tablecaption{PHOENIX Physical
  Conditions.\label{PHYSCONDTAB}} \tablehead{
  \colhead{$\tau_{std}$}&\colhead{Outflow Mass}&\colhead{Kinetic
    Energy}&\colhead{$\dot{M}$}&\colhead{Column Density}} \startdata
1&$547\ M_{\odot}$&$30\times10^{51}$
ergs&$159$~\msol~yr$^{-1}$&$2\times10^{24}$\
cm$^{-2}$\\ 10&$3000\ M_{\odot}$&$100\times10^{51}$
ergs&$466$~\msol~yr$^{-1}$&$2\times10^{25}$\ cm$^{-2}$
\enddata
\end{deluxetable}


\section{Discussion \label{Discussion}}

The model fits are especially good considering the few free parameters
in the \texttt{PHOENIX} model. The reader should keep in mind that we
are not fitting each absorption line separately, but rather construct
a global fit to the entire spectrum using  the parameters given in
Table~\ref{BESTFITTING}. This includes the absorption, the emission,
and the continuum, all fit with these few parameters.  In fact, our
model spectrum is created by the solution to the radiative transfer
equation at every wavelength point across the spectrum simultaneously.  

In this section, we discuss some of the physical constraints from
these models,  how observed polarization in FeLoBALs fits in with this
model, provide a few inferences about other FeLoBALs and then 
discuss some implications for quasar populations and evolution.

\subsection{Physical Constraints Inferred}

Although we have assumed a specific density profile, the radial
extension of our model is very small ($v_{max} = 2800$~\kmps\ and
$v_{0} = 2100$~\kmps); thus the model does not strongly probe the
density structure of the ejecta, except to infer that the density
profile is rather flat. \citet{branch02}  assumed a power-law density
profile with $\rho \propto (\frac{v}{v_0})^{-n}$, with $n=2$, whereas
our effective power-law index at the pseudo-photosphere $\tstd=1$ is
$v_0/v_e = 7$, nominally significantly steeper, but since the radial
extension of our model is so small the discrepancy should not be
considered important. What is perhaps more interesting is that our
value of $v_0$ is quite a bit higher than that of \citet{branch02} who
used a photospheric velocity of 1000~\kmps. Interestingly, they had to
impose a minimum velocity of $1800$~\kmps\ on \ion{Fe}{2} and
\ion{Cr}{2} and they chose the same value of $v_{max}=2800$~\kmps. The
lower photospheric velocity is likely to be a consequence of the 
Schuster-Schwarzschild approximation \citep{mihalas78sa} of
SYNOW. That is, in SYNOW all emission of photons occurs at the
photosphere and the atmosphere consists merely of a
``reversing-layer'' where there is no creation or destruction of
photons, only resonant scattering. In
contrast, \phx\ allows 
line formation throughout the atmosphere, and lines typically form in
the ``line-forming region'' ($3.0 < \tstd < 0.1$) depending on their
strength. The much lower ``photospheric'' temperature that we find is
robust since SYNOW does no continuum transfer and thus the
``temperature'' in a SYNOW model is not physically meaningful; it just
is a way of parameterizing the underlying continuum. However,
\phx\ solves the full NLTE radiative transfer problem and particularly
our result that the \ion{Ca}{2} H+K feature is only reproduced in NLTE
indicates that we have the right physical conditions in our model.

The synthetic model has $547 M_{\odot}$ and kinetic energy
$30\times10^{51}$ ergs above the ``photosphere''
($\tau_{std}=1$). With $R_0=1.4\times10^{17}$~cm and maximum velocity
$v_{max}=2800$~\kmps\ we estimate a crossing time of $t \simeq
\ R_0/v_{max}=15.5$~yr. We estimate the mass loss rate using $\dot{M}
\simeq M/t=35$~\msol~yr$^{-1}$ which is $1/5$  the mass loss rate in
our \phx\ model of $\dot{M}=159$~\msol~yr$^{-1}$. The kinetic energy
luminosity is estimated to be $\dot{E_k}\simeq
E_{k}/t=4.3\times10^{43}$~erg~s$^{-1}$ which is two orders of
magnitude lower than the bolometric luminosity of the model
$L_{bol}=6.3\times10^{45}$~erg~s$^{-1}$; thus the flow could be
luminosity driven. Even more interesting is that the velocity we find
is very similar to characteristic velocities of hot stellar winds
which are thought to be  driven by line absorption in the atmosphere
\citep{walborn95}.

\citet{branch02} found a luminosity of $6 \times 10^{46}$~erg~s$^{-1}$
from the photometry of \FQBS\ and using the quasar composite spectrum
of \citet{MF87} to  perform the K-correction and the relation $L_{bol}
= 9 \lambda L_\lambda$ at 5100\AA\ \citep{kaspi00b}. Using the quasar
composite spectrum of \citet{francisetal91} we find $\mathrm{L_{bol} =
  4.4 \times   10^{46}}$~erg~s$^{-1}$ for \FQBS\ and $\mathrm{L_{bol}
  = 1.7 \times   10^{45}}$~erg~s$^{-1}$ for \ISO.  Since the spectra
are so similar, but the total luminosities inferred differ by a factor
of 10, suggesting that the K-correction is important, as it is
dependent on the shape of the spectral energy distribution.

Our models have a maximum column density of
$2\times10^{25}$~cm$^{-2}$ for the entire atmosphere and a column
density of 
$2\times10^{24}$~cm$^{-2}$ for the region above the
``photosphere'' $\tau_{std}=1$. As discussed in \S~\ref{results:fqbs},
the maximum  
column density is constrained by our fit to the optical spectrum. A
higher column density would yield  higher continuum and line
optical depths and the continuum would be much weaker than
observed.  Thus, the requirement that a photosphere be present places
a limit on the minimum column density of $2\times10^{24}$~cm$^{-2}$,
and the optical spectrum places a limit on the maximum 
column density of $2\times10^{25}$~cm$^{-2}$ of the wind.
These constraints make these models Thompson thick, and is sufficient
to make these objects appear to be X-ray faint as observed
\citep{green01,gallagher06}.

We can now calculate a luminosity distance using a variant of the
Spectral-fitting Expanding Atmosphere Method (SEAM) used to derive
distances to supernovae
\citep{b93j3,l94d01,b94i2,mitchetal87a02,bsn99em04}. SEAM is a
sophisticated variant of the classical Baade-Wesselink method
\citep{baadeepm}. Our approach here differs from the conventional SEAM
method in the crucial respect that in the supernova case, we know both
that homology is an excellent approximation shortly after the
explosion and we know that there was an explosive ejection at a time
$t_0$ (which SEAM determines). Here, homology has been taken to be an
expedient Ansatz and $R_0$ has been set semi-empirically. 
There is no reason
\emph{a priori} to expect that FeLoBALs are the result of a single
ejection event and thus the radius in our models is much more poorly
known than in the case of supernovae. However the radial extent does
determine the overall density, and if we were able to identify
features which are good density indicators we may be able to place the
results of this method on firmer ground. Of course, we are sensitive
to systematic errors in the overall flux calibration and in the total
reddening. Errors due to reddening tend to cancel out, since the more
we must deredden the observed spectrum, the hotter we must make the
synthetic spectrum, which compensates for the dimmer observed
spectrum. Thus, below we will test the value used for $R_0$ by using
the synthetic spectra to  calculate synthetic photometry and the
predicted bolometric magnitudes in a number of bands. Comparing these
magnitudes to the observed photometry, we obtain a distance.  This result is
sensitive to $R_0$ and thus if we find reasonable values for the
distance we can have confidence in our choice of $R_0$, and hence that
we have the right density. The outflow rate we find is then also
reasonable. Thus, this provides an important check on our
parameters. 

The SEAM method uses observed photometry and the synthetic spectrum to
calculate synthetic photometry as well as K-corrections. From this we
find a distance modulus $\mu$ which is simply related to the
luminosity distance. Using $m_B = 17.06$ for \FQBS\ we find $\mu =
42.56$ or $d_L = 3.25$~Gpc, which compares favorably with the
luminosity distance inferred for our adopted cosmology
($\mathrm{H_0=70,\ \Omega_M=0.3,\ \Omega_{\Lambda}=0.7}$) of $d_L =
4.2$~Gpc. Using $m_B = 22.74$ for \ISO\ we find $\mu = 46.59$ or $d_L
= 20.8$~Gpc, which is a bit high compared with the luminosity distance
inferred for our adopted cosmology $d_L = 13.4$~Gpc. As we noted
above, it seems likely that the K-corrections are important and thus
we obtain a distance for the relatively nearby \FQBS\ which is good to
about 30\%, but for the more distant \ISO\ the distance is off by
55\%. Nevertheless, that fact that the distances agree to better than
a factor of two indicates that our model predicts roughly the right
size as well as the right SED for both objects.

Since our \phx\ synthetic spectra are quite a good fit, we may also
determine the bolometric luminosity using our synthetic spectra to
perform synthetic photometry and calculate K-corrections. With this
method we find that $L_{bol} = 1.2 \times 10^{46}$~erg~s$^{-1}$ for
\FQBS\ and $L_{bol} = 2.9 \times 10^{45}$~erg~s$^{-1}$ for \ISO.

We obtained $L_{bol}$ in two different ways, both of which were
consistent with each other. However the two objects have nearly
identical SEDs in the regions that we can observe
(Fig.~\ref{FQBSISO}). Why does the 
synthetic $L_{bol}$ differ by a factor of $10$? The values for $R_0$
which are the same for the synthetic models must actually differ
slightly between the two objects, or, as noted above, the K-correction
is important.

We use our model luminosity to estimate a few quantities. If we assume
that our model luminosity is the Eddington luminosity then the black
hole mass and the accretion rate are $M_E = 5 \times
10^{7}$~\msol\ and $\dot{M_E} = 1.1$~\msol~yr$^{-1}$.  These are
roughly consistent with usual estimates for quasars,  even if we scale
the luminosity up by a factor of 3 that we infer from photometry.

\subsection{Polarization\label{sec:polar}}

\citeauthor{lamy04} (\citeyear{lamy04,lamy00}) found that the polarization 
increases in the absorption troughs do not rule out the model described in
this paper. They constructed a
``two-component'' wind model.
In their model, the broad absorption
occurs in a dense equatorial wind emerging from the accretion disk,
while scattering and polarization mainly take place in a polar
region. Our model is consistent with the two component
model \citep{lamy04} in which the observer looks through a
equilateral wind  and sees a polar component which is dominated by
electron scattering.  A spherically symmetric model is a required
computational constraint inherent in \phx; thus we assume 100\%
global covering.  The polarization results indicate that some
asymmetry must be present; nevertheless, the presence of P-Cygni
profiles where the absorption troughs go almost to zero flux
indicate that the covering fraction is high. 

Our model will not change much if we relax the 100\% global covering
fraction and  have the same electron scattering polar component
described in \citet{lamy04}.  In fact, reducing the covering factor
would most likely provide a better fit, as it would reduce the
emission feature in \ion{Mg}{2} that was shown in \S 4.1 to be slightly
too big.

\subsection{Other FeLoBAL QSOs}

FeLoBALs are observed to  have a wide range of spectra \citep{hall02}.
We fit only 
the spectra of two objects in this paper; in this section, we comment
briefly on whether or not the \phx\ model is likely to be able to
explain the spectra of other objects.  

\citet{hall07} report blueshifted broad absorption lines troughs in
Balmer lines in the quasar SDSS~J125942.80+121312.6.  Our \phx\ model
predicts Balmer absorption in at least H$\beta$ and H$\gamma$ as seen
in Fig.\ 2.    Thus our model may very well be able to explain the
spectra of this object too.  

The value of the  global covering fraction is very likely to be
responsible for the differences between the spectra that we have
modelled here, and those of other  FeLoBAL QSOs
\citep{hazard87,cowie94,becker97,hall02}. Specifically, many FeLoBALs
appear to have only absorption features, and the emission features are
week.  The high global covering fraction inherent in the \phx\ model
predicts rather prominent emission features and P-Cygni profiles.
Reducing the global covering while viewing the outflow from the
equatorial region would reduce the emission line strengths while
retaining the strong absorption features. 

So-called overlapping trough QSOs \citep{hall02} may also  be able to
be explained by this paradigm. We investigated them with our models and
yet where unable to come up with a satisfactory solution, because the
emission in our 100\% global covering method is too high. Therefore
the covering fraction in these type of quasars might be  not quite as
large as for the objects modeled here.   In addition they appear to
have higher velocity winds and possibly higher optical depths.

If the covering fraction is not 100\%, then from some lines of sight
we may see the nucleus directly. The strong \ion{Fe}{2} absorption 
observed when the outflow is in the line of sight may be seen as as
emission in this case, and ultrastrong \ion{Fe}{2} emission may be
seen. The idea of linking \ion{Fe}{2} emission with BAL outflows not
in the line of sight has been suggested for the prototypical
Narrow-line Seyfert 1 (NLS1) I Zw 1 \citep{boroson92b}.

\subsection{Implications for Quasar Populations and
  Evolution \label{BAL}} 

Broad absorption lines are found in the spectra of 10--20\% of the
quasars in the Sloan Digital Sky Survey \citep[e.g.][]{trump06}.  The simplest
interpretation of this fact is that all quasars have an outflow that
occults 10--20\% of quasar lines of sight (the orientation model).
Alternatively, the outflow may cover a larger percentage of sight
lines, and the broad absorption lines may only be present during some
small fraction of the quasar lifetime during which it is blowing gas
out of the nucleus \citep[e.g.,][]{voit93, becker00, gregg06}.  As
discussed in \S~\ref{sec:polar}, our model requires a large global
covering fraction, although it does not have to be 100\%.  Here we
briefly review the support for both scenarios, and discuss how  our
results fit into models for quasars in general.

\citet{weymann91} compare the emission-line and continuum properties
of spectra from 42 BALQSOs with those of 29 normal QSOs.  They find
that the emission-line and continuum properties are very similar
between the BALQSOs and the non-BALQSOs.  This result forms a key
support for the orientation model.   Another important piece of
evidence that supports the idea that BALQSOs and non-BALQSOs differ
only in their orientation is the fact that X-ray spectra from some
BALQSOs are highly absorbed, but apparently intrinsically identical to
those of non-BALQSOs \citep{gallagher02}.

However, the \citet{weymann91} sample is very small, and the number of
BALQSOs for which X-ray spectra of sufficient quality to obtain
absorption column information is also small.  Thus, while the
orientation model is widely accepted, and may be applicable to many
BALQSOs, the evidence for the evolutionary model, at least for some
subclasses of BALQSOs, is growing.  The best candidate for objects
characterizing the evolutionary model and not fitting into the
orientation model are the low-ionization BALQSOs.  Early on, these
were noted to have optical and UV emission-line and continuum
properties different than ordinary quasars
\citep{weymann91,boroson92b}; for example, they tend to have strong
\ion{Fe}{2} 
emission and weak \ion{O}{3}]. They also tend to be more reddened than
non-BALQSOs and HiBALs \citep{reichard03}, and they are uniformly more
X-ray weak than HiBALs \citep{gallagher06}.  In addition, they are
very rare; \citet{trump06} find only about 1\% of quasars are LoBALs.
Thus, the usual argument used in favor of the orientation model, the
similarity of spectra from BALQSOs and non-BALQSOs, doesn't work as
well for LoBALs.

Further evidence for an evolutionary role of BALQSOs comes from their
radio properties.   If BALQSOs were observed predominately edge-on, as
are radio galaxies, one would expect to see a steep radio spectrum
dominated by the synchrotron emission in the radio lobes.  However,
BALQSOs show both steep and flat radio spectra
\citep[e.g.,][]{becker00}.  Furthermore, Fe II BALQSOs are extremely
rare, and there is evidence among the small sample of an
anti-correlation between the radio-loudness and the strength of the
BAL features that led \citet{gregg06} to propose that quasars in this
state are emerging from cocoons of gas that produces the BALs and
which suppresses the development of radio jets and lobes.  In
addition, \citet{brotherton06} present spectropolometric results
showing polarization parallel to the radio axis, implying a small
angle of inclination; they present an extensive review of the
implications of the radio properties on BALQSO models.

Numerous models of quasar evolution admit a time period early in the
life of a quasar when it is heavily shrouded by dust and gas.  Before
we see the bare quasar, we expect to see it in a heavily absorbed
stage.  Outflows may result from the turning on of the QSO
\citep{hazard84}.  LoBALs may be young quasars that are casting off
their cocoons of dust and gas \citep{voit93} and may be related to
ultra-luminous infrared galaxies \citep{egami99}.  \citet{becker97}
suggest that FeLoBALs may be the missing link between galaxies and
quasars.  Recent simulations of mergers and quasar evolution show that
during much of the lifetime of the quasar, it is heavily shrouded by
gas with large column densities \citep[e.g.,][]{hopkins05}.
Observational support for this view comes from the discovery of a
large number (four out of a sample of eight) LoBALs in high redshift
quasars \citep{maiolino04}.

Thus, observational evidence and scenarios for quasar evolution imply
that FeLoBALs may be candidates for support of the evolutionary
scenario.  If the quasar observable lifetime is assumed to be $10^7$
years, the fact that FeLoBALs only make up 0.33\% of quasars in the
SDSS \citep{trump06} implies that the time scale for this stage is
approximately $3 \times 10^4$ years.  That should be a lower limit,
however, since these objects may be missed in the SDSS
\citep{trump06}.  This relatively short time period, compared with the
total lifetime of a quasar,  compares favorably with our crossing time
of 15 years.

It is clear that a high global covering fraction is compatible with
the evolutionary scenario.  But what about the other parameters
obtained from the model fits?.  One possible problem is that the
radius of the model shows that the outflow occurs well within the
central engine, at approximately $1.4 \times 10^{17}\rm \,cm$.  The
quasar feedback scenario proposed by \citet{fabian99} infers that the
bulk of the gas expelled from the AGN is accelerated beyond the Bondi
radius, which for a $5 \times 10^7\,M_\odot$ black hole is $4 \times
10^{21}\rm \, cm$.  On the other hand, the amount of gas expelled
during this process could be comparable to the amount of gas contained
by the  entire galaxy, far more than the rather modest $547 M_\odot$
that we infer.  It is quite possible that the accretion/blow out
process is messy and chaotic; there may be a phase in which the
central engine ``burps'', and this results in the features that we
see.

\section{Conclusions}

Using the spectral synthesis code  \texttt{PHOENIX} we compute
synthetic spectra that provide very good fits to the observed
restframe UV and optical spectra of two FeLoBALs.
While our models are limited to exact spherical symmetry, they provide
excellent fits. In order to reconcile our results with the
polarization data on these objects, we require some asymmetry, but
still a high global covering fraction, which would only modestly
affect the flux spectrum.

We are able to determine a luminosity distance estimate which is
direct and is accurate to around 50\%. The question arises: could
these objects be used as distance indicators at high z, even if only
as a sanity check on the really high-z Hubble diagram from GRBs?

Our results lend support to the inference that FeLoBALs are an
evolutionary stage of the QSO as opposed to a pure orientation
effect. Our model with a smaller covering factor may be able to
explain other BAL QSO such as overlapping trough QSOs.  In addition
our model column densities, which are Compton thick, match those that
are  expected from X-ray non-detections of these objects.

For future work we plan to explore metallicity effects on the model
spectrum. This is not trivial, as it is not just a matter of changing
the metallicity and comparing the model. Completely different sets of
dynamical and luminosity parameters may be required to  achieve the
best fitting model spectrum. 
We plan to continue our analysis with the \citet{hall07} spectrum,
which shows $H\beta$ absorption.
We have compared our observed spectrum with the one from that paper
and think that it is an excellent candidate for this type of modeling.

\begin{acknowledgments}
  We thank David Jeffery and Sebastian Knop for helpful discussions.
  We thank Pierre-Alain Duc for the \ISO\ spectrum and Mike Eracleous
  for the optical spectrum of \FQBS.  This work was supported in part
  by NASA grants NAG5-3505, NAG5-12127, NAG5-10365 and NSF grants
  AST-0204771, AST-0506028, and AST-0307323. This research used
  resources of the National Energy Research Scientific Computing
  Center (NERSC), which is supported by the Office of Science of the
  U.S.  Department of Energy under Contract No. DE-AC03-76SF00098.  We
  thank them for a generous allocation of computer time. D.J. was
  supported in part by Project number 146001 financed by the Ministry
  of Science and Environmental Protection of Serbia.  This research
  has made use of the NASA/IPAC Extragalactic Database (NED) which is
  operated by the Jet Propulsion Laboratory, California Institute of
  Technology, under contract with the National Aeronautics and Space
  Administration.
\end{acknowledgments}


\clearpage

\clearpage

\clearpage

\clearpage

\clearpage

\clearpage

\clearpage
\appendix

\section{\label{PhxGlobalIter} The \texttt{PHOENIX} Code} 

\phx\ is a mature code.  Development of \phx\ began in 1990 and
continues today.  Discussions of the computational details appear in a
number of publications.   
We include a short description of some of the
computational details for those who are interested in understanding in
more detail how the code works.  This section also includes relevant
references.  Finally, we present a flowchart of a \phx\ computation.

\subsection{An introduction to \phx}

\phx\ is able to model astrophysical plasmas under a variety of
conditions, including differential expansion at relativistic
velocities
\citep{hb06,hb04,bh04,hbjcam99,hbapara97,phhnovetal97,ahscsarev97}. \phx\
includes very detailed model atoms constructed from the work of Kurucz
\citep{hbfe295,bhmasi97,short99} for a number of almost all the
important species (H, He, CNO, Si, Fe, Co, etc.). In addition, the
CHIANTI or APED databases may be chosen for model atoms at runtime.
The code is optimized and parallelized to run on all available
supercomputers.  \phx\ has a long history of modeling astrophysical
objects including extra-solar giant planets (EGPs), Brown dwarfs
\citep{ahscsarev97,cond-dusty}, novae \citep{petz05}, as well as all
types of supernovae
\citep{b93j3,b94i2,nughydro97,bsn99em00,l94d01,l84a01,l98s01,mitchetal87a01,bbbh06}.
In version 14, we solve the fully relativistic radiative transport
equation for a variety of spatial boundary conditions in both
spherical and plane-parallel geometries for both continuum and line
radiation simultaneously and self-consistently using an operator
splitting technique. We also use an operator splitting technique to
solve the full multi-level NLTE transfer and rate equations for a
large number of atomic species (with a total of more than 10,000
energy levels and more than 100,000 primary NLTE lines), including
non-thermal processes. MPI and OpenMP directives are used, so the code
runs on both distributed and shared memory architectures
\citep{bpar03,hlb01,bhpar298,hbapara97}.  \texttt{PHOENIX} accurately
solves the fully relativistic radiation transport equation along with
the non-LTE rate equations (currently for $\sim 150$ ions) while
ensuring radiative equilibrium (energy conservation).  Typically each
atom has several ionic species in NLTE and is represented by dozens to
hundreds of levels for the Fe-group species. \texttt{PHOENIX} is
currently around 700,000 lines of code which relies on 0.6 GB of
atomic data and 12 GB of molecular data.

In the present paper, the multilevel, non-LTE rate equations are solved
self-consistently for H I, He I--II, Mg I--III, Ca I-III, and Fe
I--III  using an accelerated lambda iteration (ALI) method
\citep{rybhum91,phhs392,hbjcam99,hbmathgesel04}. Simultaneously we
solve for the special relativistic condition of radiative equilibrium
\citep{nughydro97} using a modified Uns\"old-Lucy temperature
correction scheme.  Relativistic effects, in particular the effects of
advection and aberration, are important in the high velocity flows
observed in these quasars.

The generalized non-LTE equation of state (EOS) is solved for 40
elements and up to 26 ionization stages per element for a total of
hundreds of species.  For the conditions present in the models,
molecules are unimportant, and we neglect them in order reap
substantial savings in CPU time.  Negative ions are always
included. The numerical solution of the EOS is based on Brent's method
for the solution of nonlinear equations which is very robust and fast.

In addition to the non-LTE lines, the models include,
self-consistently, line blanketing of the most important ($\approx
10^6$) lines selected from the latest atomic and ionic line list of
Kurucz. The entire list contains close to 42 million lines but not all
of them are important for the case at hand. Therefore, before every
temperature iteration, a smaller list is formed from the original
list. A set of optical depths in the line-forming region of the gas is
chosen, then using the density and temperature for these depths, the
absorption coefficient in the line center, $\kappa_l$, is calculated
for every line and compared to the corresponding continuum (LTE+NLTE)
absorption coefficient, $\kappa_c$. A line is transferred to the
``small list'' if the ratio $\kappa_l/\kappa_c$ is larger than a
pre-specified value (in these calculations $5\times10^{-6}$, selecting
over half a million lines).  In the subsequent radiative transfer
calculations all lines selected in this way are taken into account as
individual lines and all others from the large line list are
neglected. This selection procedure is repeated at each iteration
where the pressure or temperature changes by a prescribed amount in
order to always include the most important lines. We treat line
scattering in these LTE lines by setting the albedo for single
scattering, $\alpha = 0.95$.

\subsection{A \phx\ Flowchart}

This section gives the reader a brief explanation of how the
\phx\ code computes a model spectrum. For a more complete
understanding the authors recommend the reader peruse
\citet{hbjcam99}.  Our iteration scheme for the solution of the
multi-level non-LTE problem can be summarized as follows: (1) for
given  population levels [$n_i$] and  electron densities [$n_e$], solve the
radiative transfer equation at each wavelength point and update the
radiative rates and the approximate rate operator, (2) solve the
linear system for the atomic level populations for a given electron
density, (3) compute new electron densities 
by  the generalized partition function method,  (4) calculate the
temperature corrections needed to bring the current iteration into
radiative equilibrium, (5) repeat until a fixed number of iterations
is reached (and check that all quantities have converged). It is crucial
to account for coherent scattering processes during the solution of
the wavelength dependent radiative transfer equation, it explicitly
removes a global coupling from the iterations.

As the first step in our outermost iteration loop (the model
iteration) we use the current best guess of [T, $n_i$] as function of
radius to solve the hydrostatic or hydrodynamic equations to calculate
an improved run of P$_{gas}$ with radius.  Simultaneously, the population
numbers are updated to account for changes in P$_{gas}$. The next major
step is the computation of the radiation field for each wavelength
point (the wavelength loop), which has the prerequisite of a spectral
line selection procedure for LTE background lines. Immediately after
the radiation field at any given wavelength is known, the radiative
rates and the rate operators are updated so that their calculation is
finished after the last wavelength point. In the next steps, the
population numbers are updated by solving the rate equations for each
NLTE species and new electron densities are computed, this gives
improved estimates for [$n_i$]. The last part of the model iteration is
the temperature correction scheme outlined above (using opacity
averages etc. that were computed in the wavelength loop) which
delivers an improved temperature structure. If the errors in the
constraint equations are larger than a prescribed accuracy, the
improved [T, $n_i$] are used in another model iteration. Using this
scheme, about 10-20 model iterations are typically required to reach
convergence to better than about 1\% relative errors, depending on the
quality of the initial guess of the independent variables and the
complexity of the model.


\begin{figure}
\includegraphics[height=0.9\textheight]{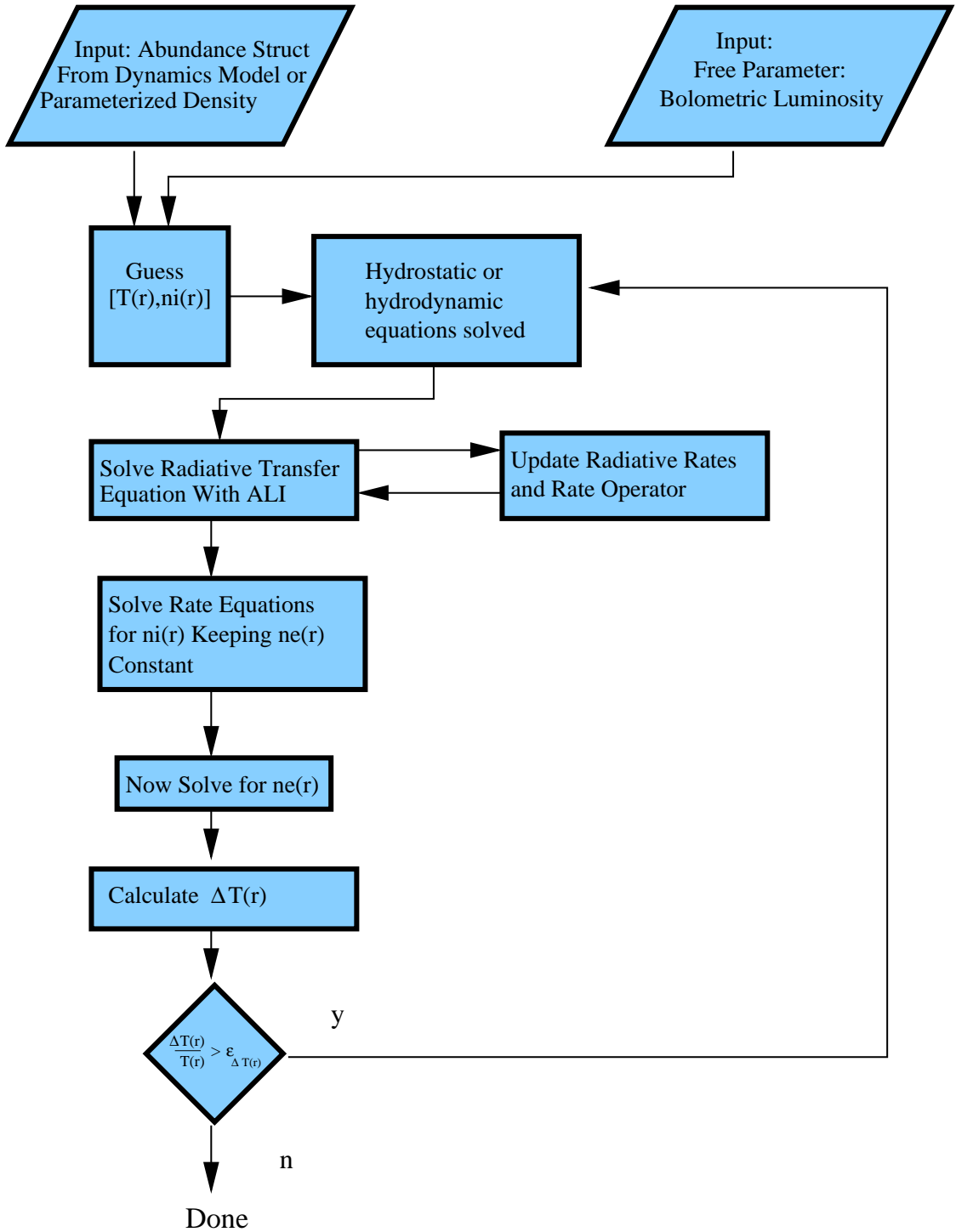}
\caption{\label{PHXGlobal} Flowchart for a global iteration of
  \texttt{PHOENIX}.  }
\end{figure}

\end{document}